\documentstyle[12pt]{article}

\parskip=\medskipamount            
\def\7#1#2{\mathop{\null#2}\limits^{#1}}        
\def\ast{\displaystyle *}

\def\beee{\begin{equation}}
\def\eeee{\end{equation}}
\def\dggg{^{\dagger}}
\begin{document}

\bibliographystyle{unsrt}

\begin{center}
{\bf QUONS IN RELATIVISTIC THEORIES\\ 
MUST BE BOSONS OR FERMIONS}\\[.1in]

Chi-Keung Chow\footnote{supported in part by the U.S.~Department of Energy
grant DE-FG02-93ER-40762;\\  email address, ckchow@physics.umd.edu.}\\
Department of Physics \\
University of Maryland\\
College Park, MD~~20742-4111\\

and\\

O.W. Greenberg\footnote{email address, owgreen@physics.umd.edu.}\\
Center for Theoretical Physics\\
Department of Physics \\
University of Maryland\\
College Park, MD~~20742-4111\\

UMPP 01-025\\
DOE/ER/40762-212\\

\end{center}

\vspace{5mm}

\begin{abstract}
The quon algebra describes particles, ``quons,'' that are
neither fermions nor bosons using a label $q$ that
parametrizes a smooth interpolation between bosons ($q = +1$) and 
fermions ($q = -1$).  We derive ``conservation of statistics''
relations for quons in relativistic theories, and show that 
in relativistic theories quons must be either bosons or fermions.
\end{abstract}

There are three reasons to study
theories that allow violations of statistics, i.e., that allow
particles that are neither bosons nor fermions.  One, which may
seem frivolous, is to stretch the framework of quantum physics and
to find out what possibilities open up when one does so.  The second,
quite concrete, is to respond to experimental interest in 
high-precision tests of the symmetrization postulate 
(that all identical particles
occur in one-dimensional representations of the symmetric group)
and of the the connection of spin and statistics (that integer-spin
particles must be bosons and odd-half-integer spin particles must be 
fermions) by providing a theoretical framework to describe such
violations together with a parameter (or parameters)
that can characterize violations if they are found
and allow quantitative bounds as well as 
the comparison of bounds on violations from different types 
of experiments if violations are not found.  
The third is to see how theories can exclude all
possibilities except bosons and fermions and thus provide an 
understanding of the empirical fact that only bosons and fermions have
been observed.

In this paper, we will only consider quantum
field theories in which the annihilation and creation operators are
constrained by a bilinear operator algebra.  This includes bosons and
fermions, and, from the point of view of the Green ansatz\cite{gre}, 
also parabosons and parafermions\cite{gre,owgmes,del} as well as 
quons\cite{owg-q=0,owg-q}.   
Since parabosons and
parafermions correspond to bosons and fermions with an exact hidden 
degree of freedom\cite{owgzwa,dhr,haag}, 
the only case that may be of interest from the 
standpoint of small violations of statistics is that of quons.  We
point out that the bilinear operator algebras that we consider give
examples of each of the possible types of identical particle
statistics in three space dimensions found in the general analysis of Haag 
and collaborators\cite{dhr,haag}.

Any theory that violates statistics must also violate one or more of
the usual conditions of relativistic quantum field theory, 
otherwise one could prove that
the symmetrization postulate and the spin-statistics connection would
hold.  The condition that quons violate is locality of observables
(which, in general, is not the same as locality in the sense of having field 
products at a single point);
however this is not tested directly to high precision.  

Let us briefly review what is known about quonic theories.  
Quon statistics is compatible with Lorentz invariance and the $CPT$ 
theorem, at least for free quon fields\cite{owg-q}.
Nonrelativistic quon theories
are valid quantum field theories (with positive squared norms) for 
$-1 \leq q \leq 1$ both for free quons and for quons that interact 
with
particle conserving interactions\cite{owgphysica}.  Bound states of 
$n$ 
quons with parameter $q$ have the parameter $q^{n^2}$\cite{owghil2}, 
a generalization of the Wigner--Ehrenfest-Oppenheimer 
result for bosons and fermions\cite{wig, ehr}.
In the nonrelativistic context, the quon theory has been used to 
parametrize
bounds on violations of statistics\cite{statexp}.
For example, as a nonrelativistic description
of electrons interacting via the Coulomb interaction, the
quon theory allows the bound 
$-1 \leq q(e) \leq -1+ 3.4 \times 10^{-26}$ to be
inferred from the experiment of E. Ramberg and G.A. Snow\cite{ram},
where $q(e)$ is the $q$-parameter for electrons and, 
as described just below, $q=-1$ is the fermion limit of quons.  

The quon theory uses a bilinear algebra,
\beee
a(k) a\dggg(l) - q(a,a) a\dggg(l) a(k) = \delta(k,l),   \label{basic}
\eeee
and the usual Fock-like vacuum condition,
\beee
a(k) | \Omega \rangle = 0,
\eeee
where $|\Omega\rangle$ is the vacuum.  
The adjoint of Eq.~(\ref{basic}) implies $q(a,a) \equiv q(a)$ is real.
It is easy to see that squared norms will be positive only for
$-1 \leq q \leq 1$.  This follows from the square of the norm of
the general two-particle state,
\[ \| \psi(k_1,k_2) a\dggg(k_1) a\dggg(k_2) | \Omega \rangle ||^2 \]
\beee
=|\psi (k_1,k_2)|^2+q(a) \psi^{\ast} (k_1,k_2) \psi (k_2,k_1) 
~k_1\neq k_2,
\eeee
where repeated indices are summed over.  For $\psi$ symmetric 
(antisymmetric) the result is $(1+q(a)) |\psi (k_1,k_2)|^2$
($(1-q(a)) |\psi (k_1,k_2)|^2$), which shows that $q(a)$ must lie in 
the closed interval $(-1,1)$ as stated above.  It is more difficult 
to show that for $q$ in this range all norms are positive or zero; 
proofs of this appear in \cite{zag, spe} among other places.  The
limiting cases, $q=-1$ ($q=1$) correspond to Fermi (Bose) statistics.

We can also define relative $q$ parameters in the bilinear relation 
between
two independent sets of annihilation and creation operators by
\beee
a(k) b\dggg(l) - q(a,b) b\dggg(l) a(k) = 0.        \label{aa}
\eeee
The adjoint of this equation,
\beee
b(l) a\dggg(k) - q(a,b)^{\ast} a\dggg(k) b(l) = 0,        \label{ab}
\eeee
leads to $q(b,a) = q(a,b)^{\ast}$.
The norm of 
$(a\dggg(k) b\dggg(l) + z b\dggg(l) a\dggg(k))| \Omega \rangle$,
is $|z|^2 + z^{\ast} q(a,b)^{\ast} +z q(a,b) +1$.  The minimum of 
this is
$1-|q(a,b)|^2$, so positivity of norms requires $|q(a,b)| \leq 1$,
the same condition as for $q(a)$, except that, so far, $q(a,b)$ does
not have to be real.

We will now study the constraints on this $q$ parameters due to the 
conservation of statistics.  
The key observation is that the Hamiltonian operator must be 
effectively a bosonic operator.  
This requirement follows from the condition that the contribution to 
the 
energy from subsystems that are widely spacelike separated should be 
additive \cite{owgphysica}.  
The terms appearing in the Hamiltonian are in general products
of field operators $\phi(x)$, which are themselves linear 
combinations of creation and annihilation operators, {\it i.e.}, 
$\phi_a(x) = \int d^3k/(2\omega_k) [a(k)exp(-ik\cdot x) + 
\bar{a}(k)\dggg exp(ik\cdot x)]$.  
As a result, there will always be a term in the Hamiltonian which is 
a product of only annihilation operators 
(or of only creation operators).  
For example, a trilinear interaction term $H_I=\phi_a \phi_b \phi_c$ 
contains a term proportional to $abc$ (and a term proportional to
$\bar{a}\dggg \bar{b}\dggg \bar{c}\dggg$).  
Since the Hamiltonian is bosonic, $abc$ should also be a bosonic 
operator and the relative $q$ factor 
$q(x,abc)$ with any annihilation operator $x$ must be one.
Thus 
\beee
x(k) (a(l_1)b(l_2)c(l_3))\dggg -q(x,abc) (a(l_1)b(l_2)c(l_3))
\dggg x(k) = \delta(k,l_1)(b(l_2)c(l_3))\dggg.                          \label{a}
\eeee
Equation (\ref{a}) leads to
\beee
q(a) q(a,b) q(a,c)= 1,
\eeee
and
\beee
q(\bar{a}, a) q(\bar{a}, b) q(\bar{a}, c) = 1,
\eeee
and cyclic permutations of $a$, $b$, $c$.
Since all the $q$'s must lie in the closed
unit disk this immediately implies 
\beee
|q| = 1
\eeee
for all $q$'s. 
Since $q(a)$'s must be real for all particles, the constraint $|q|=1$ 
implies that $q(a)$ can only be $\pm 1$; {\it i.e.}, the commutation 
relations of $a$ and $a^\dag$ must take the standard bosonic and fermionic 
form.  Thus
despite the original motivation of quonic statistics, namely to provide a 
smooth intrapolation between bosons ($q=1$) and ($q=-1$), we have come to 
the conclusion that {\it all quons in relativistic theories are either 
bosons or fermions.}  
This is the main result of this paper.  
In addition this result holds in any theory (relativistic or not) that has 
a term in its Hamiltonian with only creation or only annihilation 
operators.  

The usual rules of conservation of statistics can be recovered by setting 
$x(k) = a(k_1) b(k_2) c(k_3)$ in Eq.~(\ref{a}), which gives 
\beee
q(a,a)q(a,b)q(a,c)q(b,a)q(b,b)q(b,c)q(c,a)q(c,b)q(c,c)=1.  
\eeee
The constraints $q(a,b)=q(b,a)^*$ and $|q|=1$ imply $q(a,b)q(b,a)=1$.  
Hence the above equation simplifies to 
\beee
q(a)q(b)q(c)=1,
\eeee
which is a restatement of the Wigner--Ehrenfest--Oppenheimer theorem 
\cite{wig, ehr} with two possibilities for a three-particle vertex,\\
Case $A$: all three particles are bosons,\\ 
Case $B$: one particle is a boson and two are fermions.  

We can also easily show that the $q$ factor of a particle is equal to 
that of its antiparticle.  
Note that the Hamiltonian always contains a ``pair annihilation'' term, 
which is proportional to the the product of the annihilation operator 
$a$ and the annihilation operator of its antiparticle $\bar a$.  
(Both mass terms and kinetic terms fall into this category.)  
As a result, the relative $q$ factor between $a\bar a$ and any 
annihilation operator must be unity.  
Then it trivially follows that 
\beee
q(\bar a,b) =q(a,b)^{-1}= q(a,b)^*. \label{rec}
\eeee
Since all these $q$ are phases ($|q|=1$), Eq.~(\ref{rec}) is the 
mathematical statement that charge conjugation of one of the particles 
reverses the phase of the relative $q$ factors.  
Eq.~(\ref{rec}) implies the corollary $q(\bar a,\bar b)=q(a,b)$, and 
for the special case $a=b$, $q(\bar a)=q(a)$, which is the statement 
that the antiparticle of a boson (fermion) is also a boson (fermion).  

Our results that the diagonal $q$'s are plus or minus one and the 
off-diagonal $q$'s have absolute value one are general and hold for
theories with interactions of any finite degree.  Next we study the
trilinear interaction $H_I=\phi_a \phi_b \phi_c$ to constrain the
off-diagonal $q$'s further for this case.  Since $\bar a$ has the same 
statistics as $bc$, we have the following crossed condition:
\beee
x(k) (b(l_1)c(l_2))\dggg - q(x,\bar{a}) (b(l_1)c(l_2))\dggg x(k)  = 0,
                                                        \label{abc}
\eeee
and similarly, since $a$ has the same statistics as $\bar b \bar c$, 
\beee
x(k) (\bar{b}(l_1)\bar{c}(l_2))\dggg                  
- q(x,a) (\bar{b}(l_1)\bar{c}(l_2))\dggg x(k)  = 0,     \label{bbcc}  
\eeee
and again cyclic permutations.
In particular, by choosing $x(k) = \bar a(k)$ in Eq.~(\ref{abc}), 
we find 
\beee
q(\bar{a},\bar{a}) = q(a) = q(a,b)^{\ast}q(a,c)^{\ast}, 
\eeee
which relates $q(a,b)$ to $q(a,c)$.
This relation, and two others from cyclic permutations, relates all 
the off-diagonal $q$'s to each other.  Referring to the cases $A$ and
$B$ above, we find\\
Case $A$: $q(a)=q(b)=q(c)=1$ and $q(a,b)=q(b,c)=q(c,a)=exp(iQ)$.\\
Case $B$: $q(a)=-q(b)=-q(c)=1$ and $q(c,a)=q(a,b)=-q(b,c)=exp(iQ)$.\\
The phase angle $Q$ is a characteristic of the trilinear vertex in 
question and is not constrained by conservation of statistics alone.
 
However we now point out that, for this case, the phase angle $Q$ can be 
rotated away by a generalized Klein transformation\cite{kle}, so that, 
after the transformation, $q(a,b)=q(b,c)=q(c,a)=1$, for case $A$ and 
$q(c,a)=q(a,b)=-q(b,c)=1$ for case $B$.  We note that the true number 
operator for a field $x$, which is an infinite series in the annihilation 
and creation operators,
\beee
n_x(k) = x\dggg(k) x(k) 
+ \sum_t (1-q^2)^{-1} \sum_t (x^{\dagger}(t) x^{\dagger}(k) 
-q x^{\dagger}(k)x^{\dagger}(t))(x(l)x(t)-qx(t)x(l)) + \cdot \cdot \cdot.
\eeee
obeys
\beee
[n_x(k), x\dggg(l)]_-=\delta(k,l) x\dggg(l)
\eeee
so that the total number operator $N_x=\sum_k n_x(k)$ obeys
\beee
exp(i \phi N_x) x\dggg(k) = exp(i \phi) x\dggg(k) 
exp(i \phi N_x)                                           \label{k1}
\eeee
and
\beee
exp(i \phi N_x) x(k) = exp(-i \phi) x(k) exp(i \phi N_x)  \label{k2}
\eeee
and the number operators for independent fields commute.
Define the rephased operators by
\beee
a^{\prime}(k) = a(k) exp[i(\phi(a,a) N_a +\phi(a,b) N_b 
+\phi(a,c) N_c)]                                          \label{k3}
\eeee
and cyclic permutations, with the condition that the product 
$a^{\prime}b^{\prime}c^{\prime} = abc$ so that the
interaction Hamiltonian is not changed.  We also require the standard
relative commutation relations stated above.  
Straightforward calculations using Eq.~(\ref{k1}), (\ref{k2}) and (\ref{k3}) 
show that the standard forms for both cases $A$ and $B$ result when the 
phases are chosen as follows
\beee 
\phi(a,a)=\phi(b,b)=\phi(c,c)=\phi(a,b)=\phi(b,c)=\phi(c,a)=-Q/3.
\eeee
Thus for the case of a single trilinear interaction the conservation of 
statistics rules together with the generalized Klein transformations 
lead to the standard results that fields are either bosons or fermions
and the relative commutation relations are bosonic unless both fields
are fermions, in which case the relative relation is fermionic.
It is plausible that the standard results also hold for theories
with several fields and with interactions of finite higher degree.
A likely way to study this question is to generalize the technique
of H. Araki\cite{ara} which he used to demonstrate the standard form
for relative commutation relations in theories with bosons and fermions.
If the phases cannot be removed in the general case, it would be interesting
to study their physical significance.

Returning to a more phenomenological note, we apply these results to 
electrodynamics, $H_I = e^+ e^- \gamma$,
where we choose $b = e^+$, $c = e^-$ and $a = \gamma$. We find
\beee
q(e^+)=q(e^-)=q(e^+, e^-) = \pm 1,
\eeee
and
\beee
q(\gamma) = q(e^+, \gamma) = q(e^-, \gamma) = 1.
\eeee
It is not surprizing that although
we can prove the symmetrization postulate for electrons and photons
and the spin-statistics connection for photons, we cannot prove the
spin-statistics connection for electrons since our formalism does
not use local commutativity of observables (in the sense that 
observables commute at spacelike separtion) and the spin of the fields
did not enter our calculation.
In other words, we have not specified whether we are studying 
electrodynamics with spinor or scalar electrons, both of which would be 
described by the schematic interaction Hamiltonian $H_I$ given above.  

In passing we discuss the corresponding constraints for a 
nonrelativistic theory.  Since there is no crossing symmetry in a 
nonrelativistic theory, the fact that the 
process $a \rightleftharpoons bc$ is allowed does not imply that
crossed processes like $c \rightleftharpoons b \bar{a}$ can occur;
thus the constraints are weaker.  (Note that the process we consider
here is not one of the crossed processes we discussed in the 
relativistic case.)  Following the pattern of the arguments
given above we find $q(a) = q(a,b) q(a,c)$, $q(a,b) = q(b) q(b,c)$, 
and $q(a,c) = q(b,c) q(c)$.  In contrast to the relativistic case in 
which all diagonal $q$'s have magnitude one, in the nonrelativistic 
case only the weaker constraint $|q| \leq 1$ holds.  For 
nonrelativistic electrodynamics with $a=b=e^-$ and $c=\gamma$,
the constraints are $q(e^-)=q(e^-) q(e^-, \gamma)$ and 
$q(e^-, \gamma) = q(e^-,\gamma) q(\gamma)$.  Together these imply
$q(\gamma) = q(e^-, \gamma) =1$.  The electron $q$-parameter $q(e^-)$
is unconstrained.

We also note that the result $q(bound) = q^{n^2}$ for a bound state of
$n$ quons in a nonrelativistic theory
can be found using the technique just given.  If the 
constituents, $a$, obey
\beee
a(k) a\dggg(l) -  q a\dggg(l) a(k) = \delta(k,l)
\eeee
then, since the transition $a^n \rightleftharpoons b$ is allowed, 
the bound state, $b$, of $n$ $a$'s should obey
\beee
b(k) b\dggg(l) - q^{n^2} b\dggg(l) b(k) = \delta(k,l) 
+ {\rm nonleading terms},
\eeee
\beee
a(k) b\dggg(l) - q^n b\dggg(l) a(k) = 0.
\eeee
Thus as found earlier\cite{owghil2} $q(bound) = q^{n^2}$.

In this paper we studied the implications of conservation of statistics 
for theories governed by a generalized commutation relation that 
involves bilinears in the creation and annihilation operators.  
Bilinear relations are a natural type of algebra to
characterize statistics since they include the bose and fermi cases
and with the Green ansatz the parabose and parafermi
cases as well as the quons.  The demonstration that quons in 
relativistic theories must be either bosons or fermions is a step in the 
direction of understanding the experimental absence of statistics
other than bose or fermi in three-dimensional space.

\end{document}